\pdfoutput=1
\documentclass[preprint,showpacs,preprintnumbers,amsmath,amssymb,floatfix,endfloats*]{revtex4}

\usepackage{graphicx}
\usepackage{dcolumn}
\usepackage{bm}
\usepackage{amsfonts}

\DeclareMathAlphabet{\mathsfsl}{OT1}{cmr}{bx}{it}
\begin{document}
\title{Evolution of the pore size distribution in sheared binary glasses}
\author{Nikolai V. Priezjev$^{1}$ and Maxim A. Makeev$^{2}$}
\affiliation{$^{1}$Department of Mechanical and Materials
Engineering, Wright State University, Dayton, OH 45435}
\affiliation{$^{2}$Department of Chemistry, University of
Missouri-Columbia, Columbia, MO 65211}
\date{\today}
\begin{abstract}

Molecular dynamics simulations are carried out to investigate
mechanical properties and porous structure of binary glasses
subjected to steady shear.  The model vitreous systems were prepared
via thermal quench at constant volume to a temperature well below
the glass transition. The quiescent samples are characterized by a
relatively narrow pore size distribution whose mean size is larger
at lower glass densities.  We find that in the linear regime of
deformation, the shear modulus is a strong function of porosity, and
the individual pores become slightly stretched while their
structural topology remains unaffected.   By contrast, with further
increasing strain, the shear stress saturates to a density-dependent
plateau value, which is accompanied by pore coalescence and a
gradual development of a broader pore size distribution with a
discrete set of peaks at large length scales.

\end{abstract}



\maketitle

\section{Introduction}

The design and synthesis of novel porous materials with advanced
mechanical and thermal properties are important for various
structural, environmental and biomedical applications~\cite{Su17}.
In particular, a detailed understanding of the relationship between
mechanical properties, including elastic and shear moduli, and
porous structure of materials remains a major challenge.  In the
last two decades, a number of studies have numerically reconstructed
the three-dimensional microstructure in silica and ceramic
glasses~\cite{Gubbins99,Atkinson15} and metallic
foams~\cite{Brinson06,Singh09,Tsipas10}.  For example, it was
recently demonstrated that during the initial stage of sintering of
ceramic films, the elastic modulus increases with the neck size of
adjacent mono-size particles, while at the later stage of the
coarsening process, the modulus becomes insensitive to the details
of the microstructure and only depends on
porosity~\cite{Atkinson15}.  Moreover, the time evolution of the
pore size distribution during the foaming process based on growth of
pressurized pores was simulated using the data from two-dimensional
metallographic images, and the resulting microstructure was used as
an input for the finite element analysis of the stress-strain
response of porous titanium~\cite{Brinson06}.

\vskip 0.05in

In recent years, molecular dynamics (MD) simulations were used
extensively to study the structure and mechanical properties of
porous glassy
materials~\cite{Vashishta93,Makeev94,Ogata99,Foffi05,Yang10,Kob11,Kob14,Rimsza14,Foffi14,Eckert16,Makeev17}.
In particular, it was shown that upon tensile loading, a periodic
array of pores in metallic glasses causes strain localization along
planes that contain pores and thus promotes the initialization and
propagation of shear bands~\cite{Yang10,Eckert16}.   Interestingly,
under large compressive deformation, the annihilation of adjacent
pores can hinder propagation of shear bands and result in temporal
hardening of porous metallic glasses~\cite{Yang10}.  Furthermore, it
was found that in nanophase silica glasses, prepared by sintering at
different pressures, the structure of pores is self-similar with the
fractal dimension of about 2, and the short range order is similar
to bulk glasses, unlike the intermediate range order where the first
sharp diffraction peak is smaller and shifted to smaller wave
vectors~\cite{Ogata99}.   The analysis of the MD data indicated that
the elastic moduli of nanophase silica glasses scale with the
density as $\sim\rho^{3.5}$~\cite{Ogata99}, while in the other MD
study, the dependence of the elastic moduli on porosity in
nanoporous silica was fitted by either exponential or power-law
functions~\cite{Rimsza14}.   However, the exact relationship between
the elastic, shear, and bulk moduli and the pore structure and size
distribution in amorphous materials has not been established.

\vskip 0.05in

In the previous MD studies, the microtopography of binodal glassy
systems at a low temperature was meticulously analyzed in quiescent
samples of different densities~\cite{Kob11,Kob14,Makeev17}. Notably,
it was found that the average pore size depends strongly on
porosity, and the pore size distributions follow a universal scaling
relation at small and intermediate length scales~\cite{Makeev17}. In
addition, the multi-fractal features of the microtopography of the
porous systems were identified at small length scales by the
analysis based on the generalized correlation integral
method~\cite{Makeev17}. It was recently demonstrated that, under
athermal quasistatic shear deformation, the structural evolution of
porous glasses involves a formation of large voids, while the solid
network fractures and becomes more compact with increasing strain
until all atoms belong to a single large cluster~\cite{Foffi14}.
Nevertheless, the time evolution of the porous network in sheared
glassy solids at finite temperatures remains not fully understood.

\vskip 0.05in


In this paper, we study the dynamic response and evolution of
three-dimensional porous structure in steadily sheared binary
glasses using molecular dynamics simulations. The quiescent samples
are produced via a coarsening process following a thermal quench
across the glass transition at constant volume. It will be shown
that with increasing shear strain, the individual pores become
significantly deformed and in some cases connect with each other,
leading to formation of large pores that are comparable with the
system size. The results of molecular dynamics simulations indicate
the power-law dependence of shear modulus on the average glass
density.

\vskip 0.05in


The rest of the paper is organized as follows. The molecular
dynamics simulation model, the glass preparation procedure, and the
deformation protocol are described in the next section.  The
analysis of atom configurations and shear stress in quiescent and
sheared porous glasses as well as the probabilities of the pore size
distribution are presented in Sec.\,\ref{sec:Results}.  A summary of
the results is given in the last section.

\section{Molecular dynamics simulations}
\label{sec:MD_Model}

\begin{figure}[t]
\includegraphics[width=15.cm,angle=0]{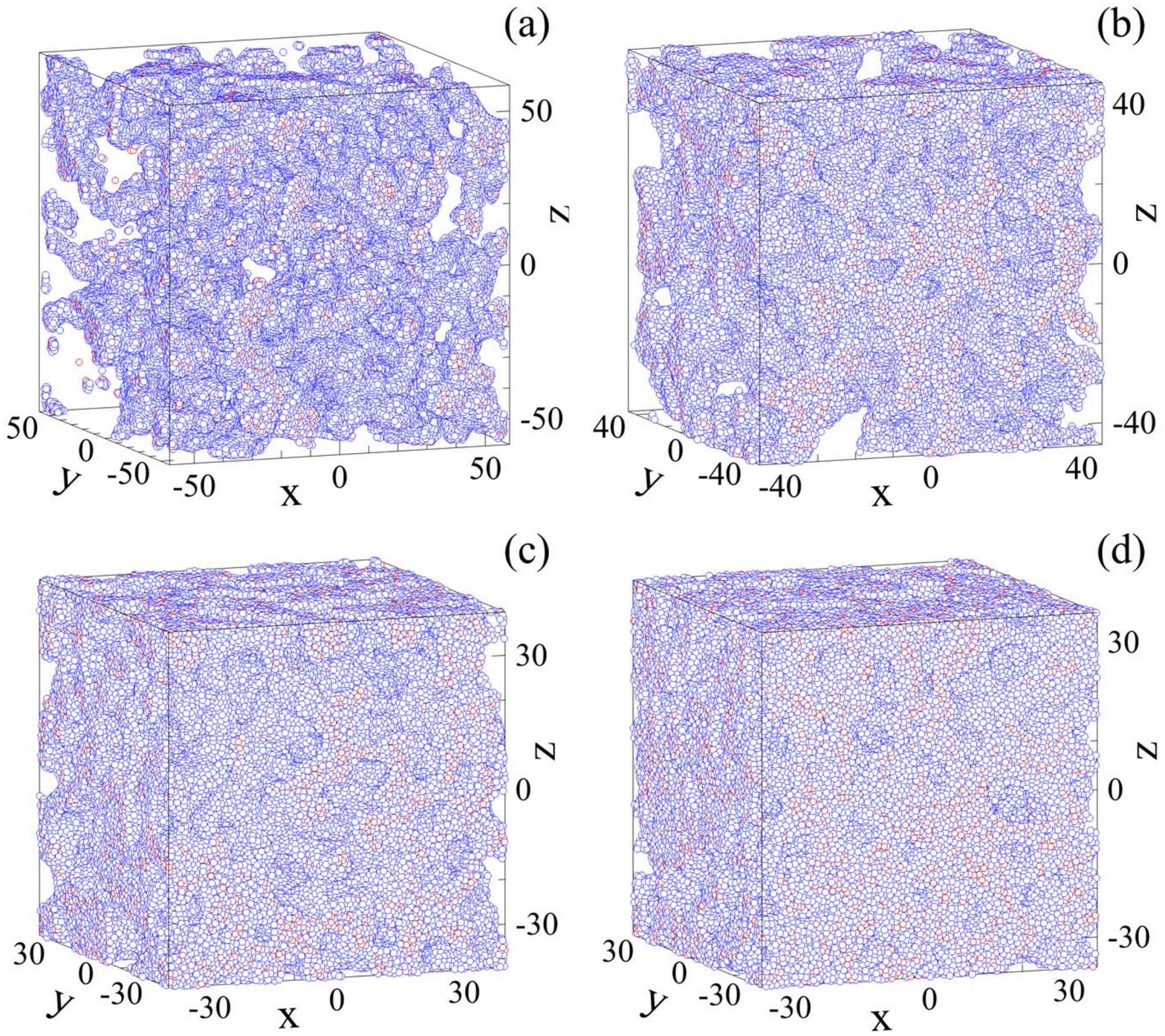}
\caption{(Color online) Instantaneous atom configurations of the
porous binary glass at the temperature $T=0.05\,\varepsilon/k_B$ for
the indicated average densities (a) $\rho\sigma^{3}=0.2$, (b)
$\rho\sigma^{3}=0.4$, (c) $\rho\sigma^{3}=0.6$, and (d)
$\rho\sigma^{3}=0.8$. Atoms of types A and B are denoted by blue and
red circles respectively. Note the different scales in all panels. }
\label{fig:snapshot_system}
\end{figure}

We study a simple glass-forming system obtained via a deep quench at
constant volume from a liquid state to a low temperature well below
the glass transition~\cite{Kob11,Kob14,Makeev17}.    During
coarsening process following the thermal quench, the system
gradually evolves into a glassy state with porous structure, as
shown in Fig.\,\ref{fig:snapshot_system}.   In the binary (80:20)
mixture model, the pairwise interaction between nearby atoms is
described by the truncated Lennard-Jones (LJ) potential:
\begin{equation}
V_{\alpha\beta}(r)=4\,\varepsilon_{\alpha\beta}\,\Big[\Big(\frac{\sigma_{\alpha\beta}}{r}\Big)^{12}\!-
\Big(\frac{\sigma_{\alpha\beta}}{r}\Big)^{6}\,\Big],
\label{Eq:LJ_KA}
\end{equation}
with the interaction parameters $\varepsilon_{AA}=1.0$,
$\varepsilon_{AB}=1.5$, $\varepsilon_{BB}=0.5$, $\sigma_{AB}=0.8$,
$\sigma_{BB}=0.88$, and $m_{A}=m_{B}$~\cite{KobAnd95}. Throughout
the study, the cutoff radius is fixed
$r_{c,\,\alpha\beta}=2.5\,\sigma_{\alpha\beta}$. All quantities are
expressed in the LJ units where length, mass, energy, and time are
measured in $\sigma=\sigma_{AA}$, $m=m_{A}$,
$\varepsilon=\varepsilon_{AA}$, and
$\tau=\sigma\sqrt{m/\varepsilon}$, respectively.   The Newton's
equations of motion were integrated using the velocity Verlet
algorithm~\cite{Allen87,Lammps} with the time step $\triangle
t_{MD}=0.005\,\tau$.

\vskip 0.05in


Initially, the atoms were placed into a cubic box and the system was
equilibrated at the high temperature of $1.5\,\varepsilon/k_B$,
where $k_B$ is the Boltzmann constant, during the time interval of
$3\times 10^4\,\tau$ at constant volume.   In the quiescent system,
the temperature was regulated by simple velocity rescaling.  It was
previously shown that the critical temperature of the binary LJ
model at $\rho\sigma^{3}=1.2$ is
$T_c\approx0.435\,\varepsilon/k_B$~\cite{KobAnd95}. In all our
simulations, the total number of atoms of type A is $N_A=240\,000$
and type B is $N_B=60\,000$. The equilibration of the liquid phase
was carried out in a wide range of densities $0.2\leq\rho\sigma^{3}
\leq 1.0$ in five independent samples for each value of the glass
density. After the equilibration procedure, the temperature was
instantaneously reduced to the target temperature of
$0.05\,\varepsilon /k_{B}$, and then the system was allowed to
evolve at constant volume for an additional time interval of
$10^{4}\,\tau$ to form a porous glass.   Typical snapshots of atom
configurations are presented in Fig.\,\ref{fig:snapshot_system} for
different average glass densities.

\vskip 0.05in


Next, the system was subjected to steady shear along the $xz$ plane
at a constant volume using the Lees-Edwards boundary
conditions~\cite{Evans90} with the strain rate of
$10^{-4}\,\tau^{-1}$ during the time interval of $10^{4}\,\tau$.
Thus, the maximum shear strain for all glass densities considered in
the present study is $100\,\%$. In the deformation protocol, the
temperature of $0.05\,\varepsilon /k_{B}$ was maintained by the
Nos\'{e}-Hoover thermostat with the coupling parameter of
$1.0$~\cite{Lammps}.  At the low temperature $T=0.05\,\varepsilon
/k_{B}$, a typical nonequilibrium simulation during $10^{4}\,\tau$
using 24 parallel processors required about $14$ hours.   During the
production run, the shear stress was computed every $0.5\,\tau$ and
the positions of all atoms were saved every $500\,\tau$ and then
used in the postprocessing analysis of the pore size distribution
and density profiles.

\section{Results}
\label{sec:Results}


The formation of disordered porous structures during spinodal
decomposition of a glass-forming liquid is influenced by a number of
factors including the rate of thermal quench, target temperature,
average density, and type of thermodynamic process~\cite{Foffi05}.
While the binary system becomes fully demixed at either high
temperatures or during slow quench, the separation kinetics is
significantly slowed down after a fast quench to temperatures well
below the glass transition, which leads to formation of structures
with complex biphasic morphologies~\cite{Kob11,Kob14}. In the
present study, the binary mixture was instantaneously brought from a
high-temperature molten state to a glassy solid well below the glass
transition. At a constant volume, the system slowly evolved towards
the final structure with porous morphology that depends crucially on
the average density. The representative atom configurations are
displayed in Fig.\,\ref{fig:snapshot_system} for the selected values
of the average glass density $\rho\sigma^{3}=0.2$, $0.4$, $0.6$ and
$0.8$ at the temperature $T=0.05\,\varepsilon /k_{B}$.  It can be
clearly seen that with decreasing glass density, the pores become
larger and more interconnected.   It was previously shown that the
probability distribution of pore sizes in quiescent samples is well
described by a Gaussian function at high densities, while as the
average glass density decreases, the shape of distributions becomes
asymmetrical and skewed toward larger length scales~\cite{Makeev17}.

\vskip 0.05in

\begin{figure}[t]
\includegraphics[width=12.cm,angle=0]{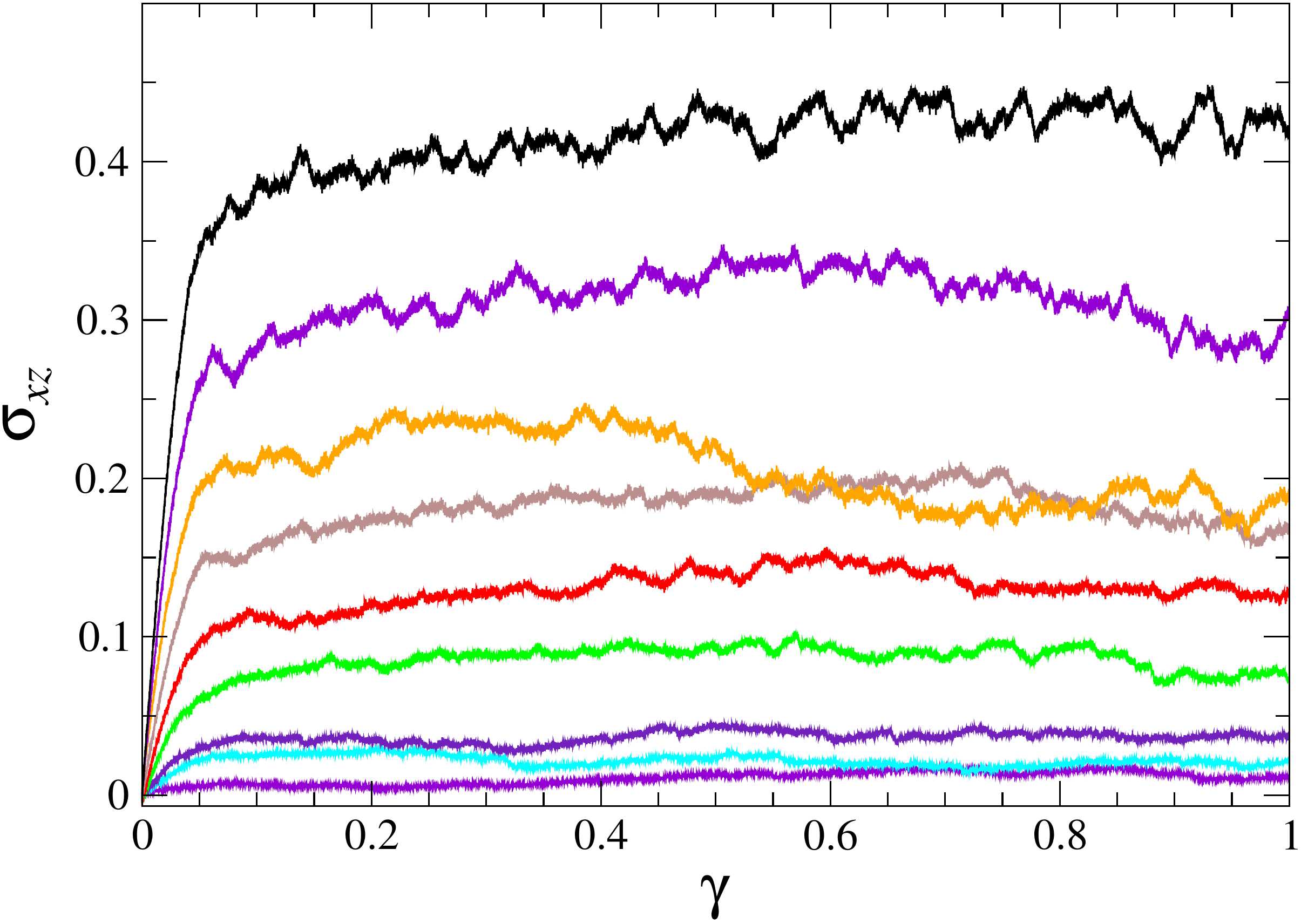}
\caption{(Color online) The variation of shear stress $\sigma_{xz}$
(in units of $\varepsilon\sigma^{-3}$) as a function of strain for
the average glass densities $\rho\sigma^{3}=0.2$, $0.3$, $0.4$,
$0.5$, $0.6$, $0.7$, $0.8$, $0.9$ and $1.0$ (from bottom to top).
The shear rate is $\dot{\gamma}=10^{-4}\,\tau^{-1}$ and temperature
is $T=0.05\,\varepsilon/k_B$. }
\label{fig:stress_strain}
\end{figure}

The dependence of shear stress as a function of strain in one sample
is presented in Fig.\,\ref{fig:stress_strain} for different
densities.    As is evident, in all cases, the stress increases
monotonically up to a few percent strain until it gradually reaches
a quasi-plateau and then decreases slightly at large strain.  Note
that the yield peak is absent in stress-strain curves even at higher
densities, which is consistent with the results for sheared
homogeneous glasses that were prepared by fast thermal
quenching~\cite{FalkPRL05}.   It can be observed from
Fig.\,\ref{fig:stress_strain} that both the initial slope and the
average value of the stress plateau become larger at higher glass
densities.   Notice, however, the presence of a significant
deviation in shear stress from its average value at large strain
$\gamma \gtrsim 0.5$ for the case $\rho\sigma^{3}=0.8$, which is
indicative of large shear-induced structural changes (discussed
below).   We further comment that results from test runs for very
large shear strain deformation ($\gamma \lesssim 5$) showed
pronounced stress oscillations with a typical period of about
$10^{4}\,\tau$ (i.e., $\gamma=1$), which implies that large solid
domains can breakup and unphysically reconnect via periodic boundary
conditions. Therefore, in the present study, shear strain
deformation only up to $\gamma \leqslant 1$ was considered.

\vskip 0.05in

\begin{figure}[t]
\includegraphics[width=12.cm,angle=0]{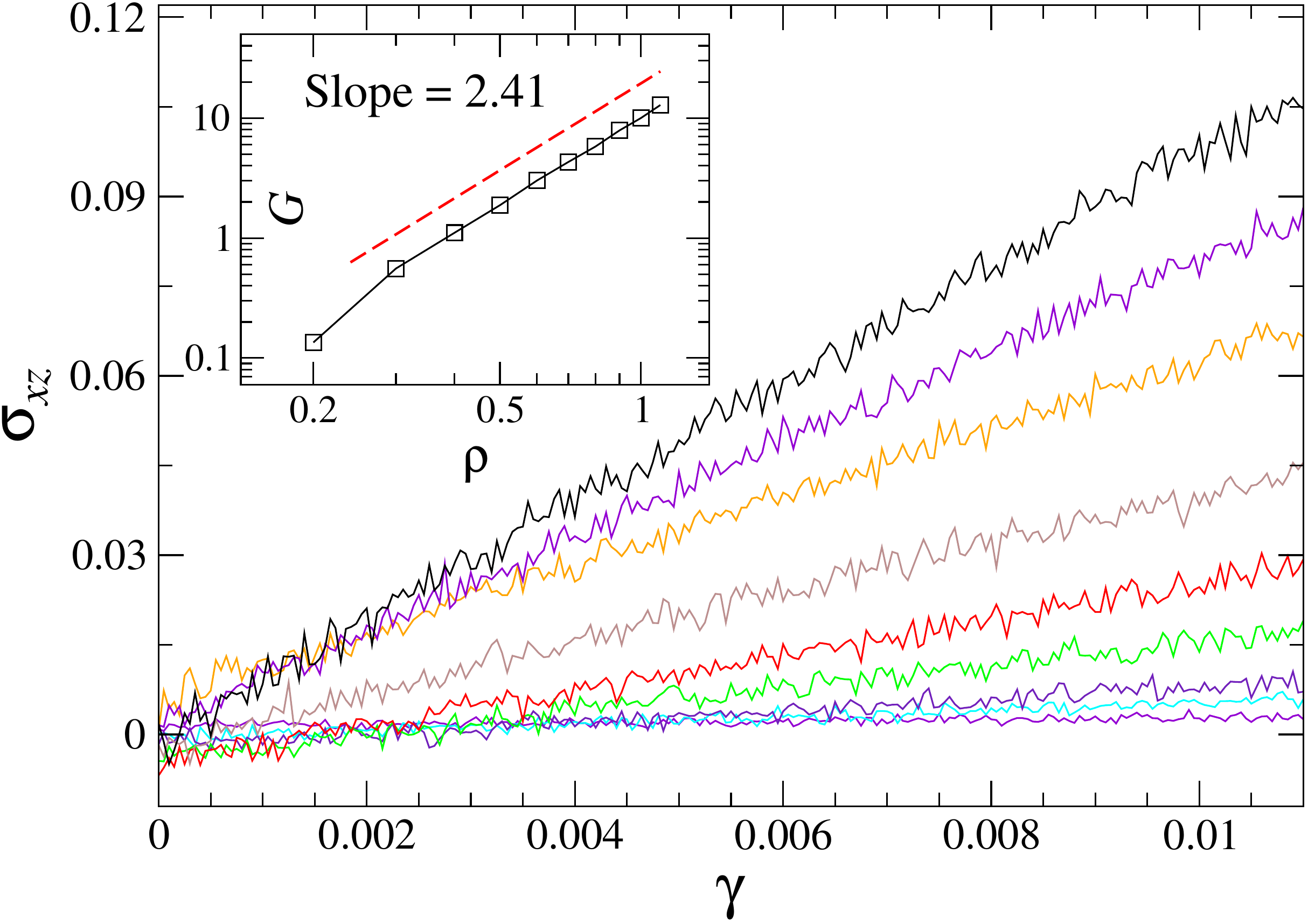}
\caption{(Color online) The shear stress-strain curves at small
strain $\gamma \lesssim 0.01$ and densities $\rho\sigma^{3}=0.2$,
$0.3$, $0.4$, $0.5$, $0.6$, $0.7$, $0.8$, $0.9$ and $1.0$ (from
bottom to top). The same data and colorcode as in
Fig.\,\ref{fig:stress_strain}. The inset shows the shear modulus $G$
(in units of $\varepsilon\sigma^{-3}$) as a function of average
glass density $\rho\sigma^{-3}$. The data were averaged over five
independent samples. The red dashed line indicates the slope of
$2.41$. }
\label{fig:stress_strain_1pr}
\end{figure}
\begin{figure}[t]
\includegraphics[width=12.cm,angle=0]{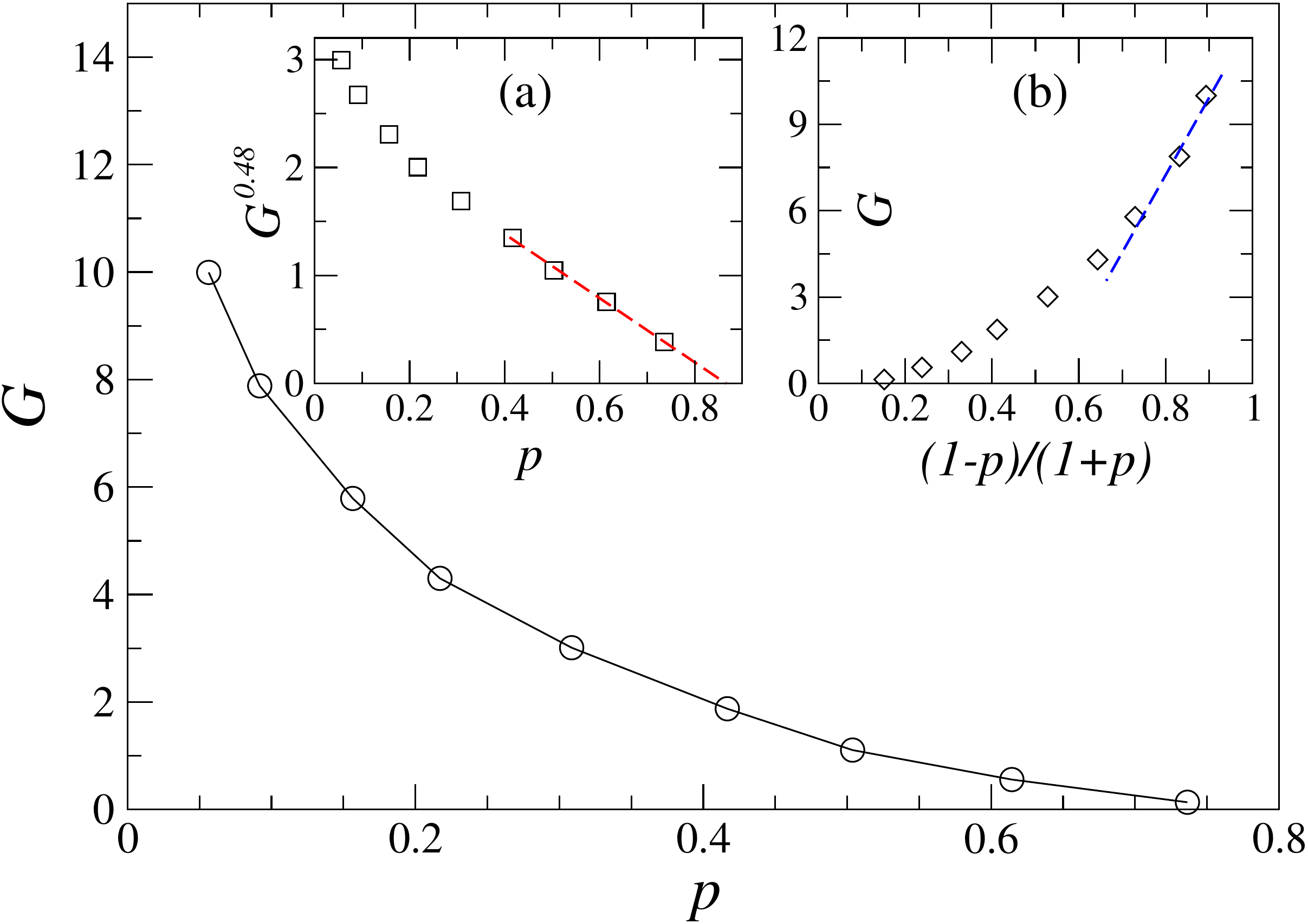}
\caption{(Color online) The variation of shear modulus $G$ (in units
of $\varepsilon\sigma^{-3}$) as a function of porosity $p$. The
inset (a) shows the same data plotted as $G^{0.48}$ vs. $p$. The red
dashed line is the best fit to the data for $p\gtrsim0.4$. The inset
(b) shows the dependence of $G$ as a function of the variable
$(1-p)/(1+p)$. The blue dashed line is a guide for the eye.}
\label{fig:shear_mod_vs_porosity}
\end{figure}

An enlarged view of stress-strain curves at small values of strain,
$\gamma \lesssim 0.01$, is displayed in
Fig.\,\ref{fig:stress_strain_1pr} for the same set of densities as
in Fig.\,\ref{fig:stress_strain}.  It is clearly seen that the shear
stress increases linearly, with superimposed fluctuations, as a
function of strain and the slope of the curves is larger at higher
densities. We also find that at very small values of strain, $\gamma
\lesssim 0.001$, the shear stress in some samples is slightly
negative, see Fig.\,\ref{fig:stress_strain_1pr}.  The observed small
deviation from zero stress in the absence of externally imposed
strain appears as a result of the coarsening process where internal
stresses become frozen in the porous
structure~\cite{Takeuchi81,Kob14}. Furthermore, in the inset to
Fig.\,\ref{fig:stress_strain_1pr}, the average shear modulus, which
was computed from the linear slope of the stress-strain data at
$\gamma \lesssim 0.01$, is shown as a function of the average glass
density. As expected, the shear modulus strongly depends on the
average density. The best fit to the MD data suggests the following
scaling with the density $G\sim\rho^{2.41}$. Further insight can be
gained by replotting the same data for the shear modulus as a
function of porosity, as illustrated in
Fig.\,\ref{fig:shear_mod_vs_porosity}.


\vskip 0.05in


In general, there exist two classes of models designed to describe
the elastic properties of porous materials. The models belonging to
the first class derive from the pioneering work by Eshelby on single
ellipsoidal heterogeneity in an elastic medium~\cite{Eshelby57}.
Since the Eshelby's work, a considerable progress has been made to
describe the mechanical response of heterogeneous systems. Several
theoretical approaches, developed to date, specifically focus on the
elastic moduli dependence on the volume fraction of voids, or the
material's macro-porosity, $p$. The major approaches include - but
not limited to: 1) the mean-field homogenization
method~\cite{Mura87,Chris91}, 2) the Mori-Tanaka
model~\cite{Mori73}, 3) the generalized self-consistent
model~\cite{Chris79}, and 4) the differential
method~\cite{Roscoe52}.   It is not counterintuitive to hypothesize,
however, that the above approaches (based on the Eshelby's
tensors~\cite{Eshelby57}) are good candidates for the problems
dealing with ensembles of non-overlapping inclusions; that is, at
relatively low porosity - well above the percolation threshold,
$p_c$.   Note that the vast majority of experimental data available
to date are obtained for porous materials with porosity below $0.5$
(see, e.g.,~\cite{Spinner63,Panakkal90,Wang84,Coronel90}).
The highly porous systems may require a different approach. Indeed,
the general topography of highly porous solids is characterized by
length scales, which can be lower than the limits of applicability
of the continuum elasticity theory. Moreover, the fractal properties
of pores alone can introduce significant corrections to any
continuum elasticity-based models.

\vskip 0.05in


The second class of models describing the elastic moduli in porous
media is based on the concept of percolation. Correspondingly, in
the frameworks of these models it is postulated that the elastic
moduli dependence on porosity can be represented by the relation
$E\sim(p_c-p)^f$, where $E$ is the elastic modulus of the material.
In three dimensions, the prediction of the percolation theory for
the Young's modulus is: $f = 2.1$~\cite{Sahimi94}. In the present
study, we used both classes of models to fit our data - that is, we
separated the regions of small porosity (ensemble of isolated
inclusions in the elastic continuum; $p < 0.5$) and
near-percolation-threshold network ($p > 0.5$). In the latter case,
we used the following functional form for the shear modulus:
$G=G_0\,(p_c-p)^f$. This approach leads to the following numerical
values of the exponent and percolation threshold: $f = 2.1$ and $p_c
= 0.87$ [\,see the inset (a) in
Fig.\,\ref{fig:shear_mod_vs_porosity}\,].  In the former case, the
dependence $G(p)$ at small values of porosity is best reproduced by
the relation $G\sim(p-1)/(p+1)$, as shown in the inset (b) to
Fig.\,\ref{fig:shear_mod_vs_porosity}.  It should be emphasized,
that the finite-size of the systems under consideration can
introduce some corrections to the aforedescribed behavior. However,
we do not expect these corrections to be significant. As was shown
in the past, the behavior follows $\sim L^{1/0.9}$
relation~\cite{Sahimi94}, which is a minor effect for the system
sizes we consider.  One additional remark can be in order. As
experimental studies have shown, the exponent for the elastic moduli
can differ substantially from the theoretical predictions.  For
example, in Ref.\,\cite{Kovacik98}, the reported value of the
power-law exponent for aluminum foam is $f = 1.65$. Thus, further
studies are of the essence.

\vskip 0.05in

\begin{figure}[t]
\includegraphics[width=15.cm,angle=0]{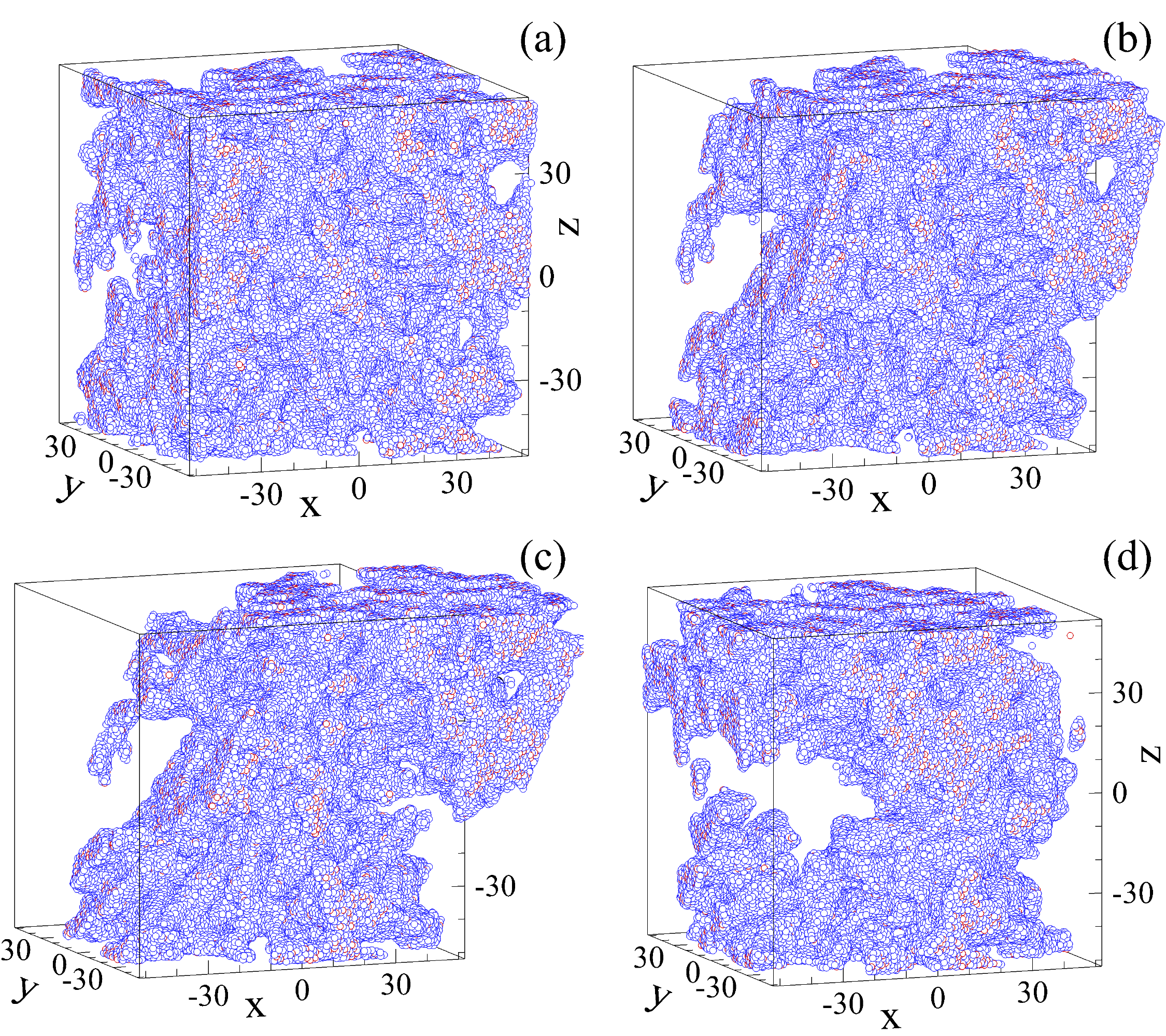}
\caption{(Color online) A sequence of atom positions for the average
glass density $\rho\sigma^{3}=0.3$ and shear strain (a)
$\gamma=0.05$, (b) $\gamma=0.25$, (c) $\gamma=0.45$, and (d)
$\gamma=0.95$. The strain rate is $10^{-4}\,\tau^{-1}$. }
\label{fig:snapshot_shear_rho03}
\end{figure}
\begin{figure}[t]
\includegraphics[width=15.cm,angle=0]{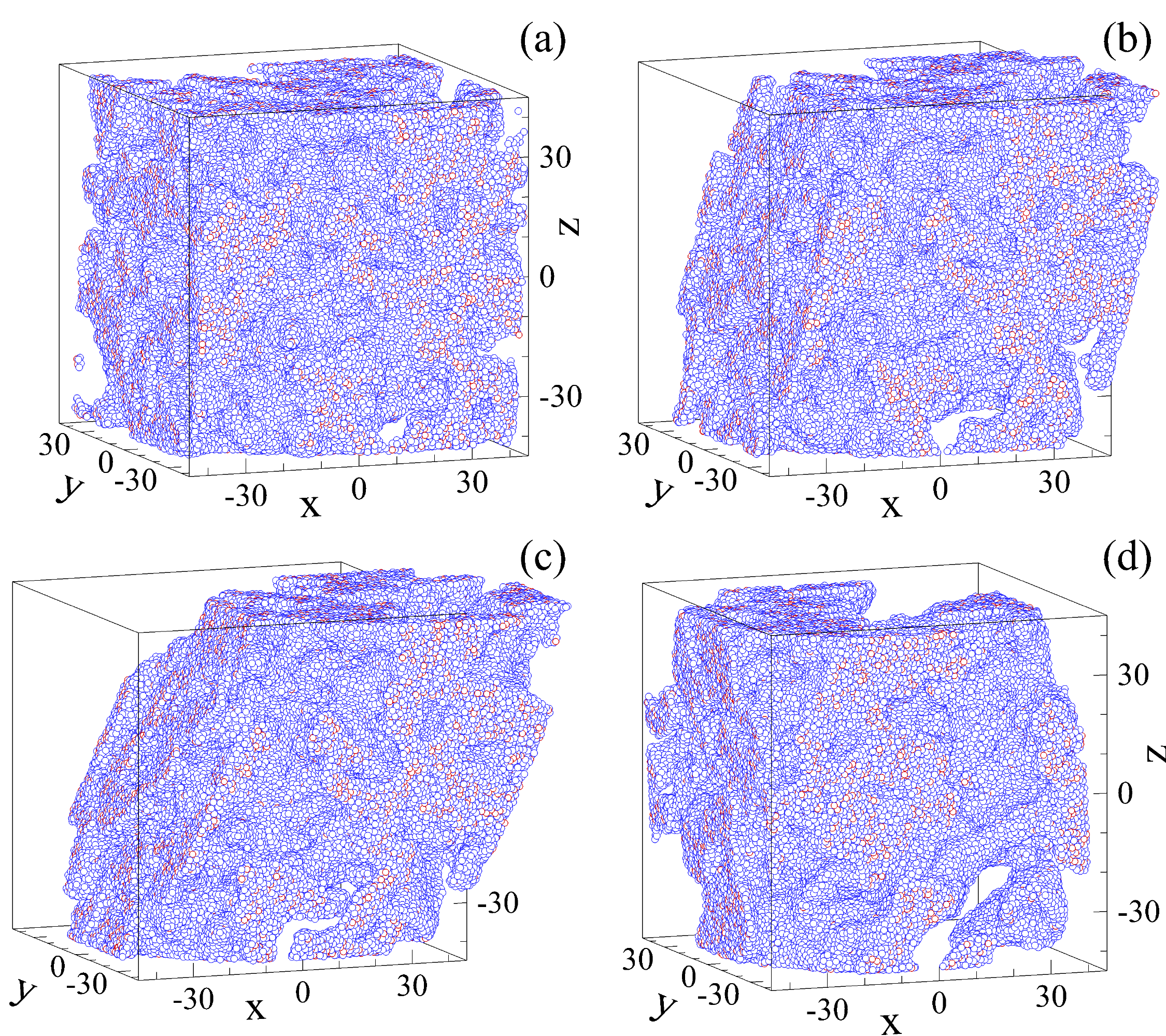}
\caption{(Color online) Snapshots of atom positions for the average
glass density $\rho\sigma^{3}=0.5$ and shear strain (a)
$\gamma=0.05$, (b) $\gamma=0.25$, (c) $\gamma=0.45$, and (d)
$\gamma=0.95$.}
\label{fig:snapshot_shear_rho05}
\end{figure}
\begin{figure}[t]
\includegraphics[width=15.cm,angle=0]{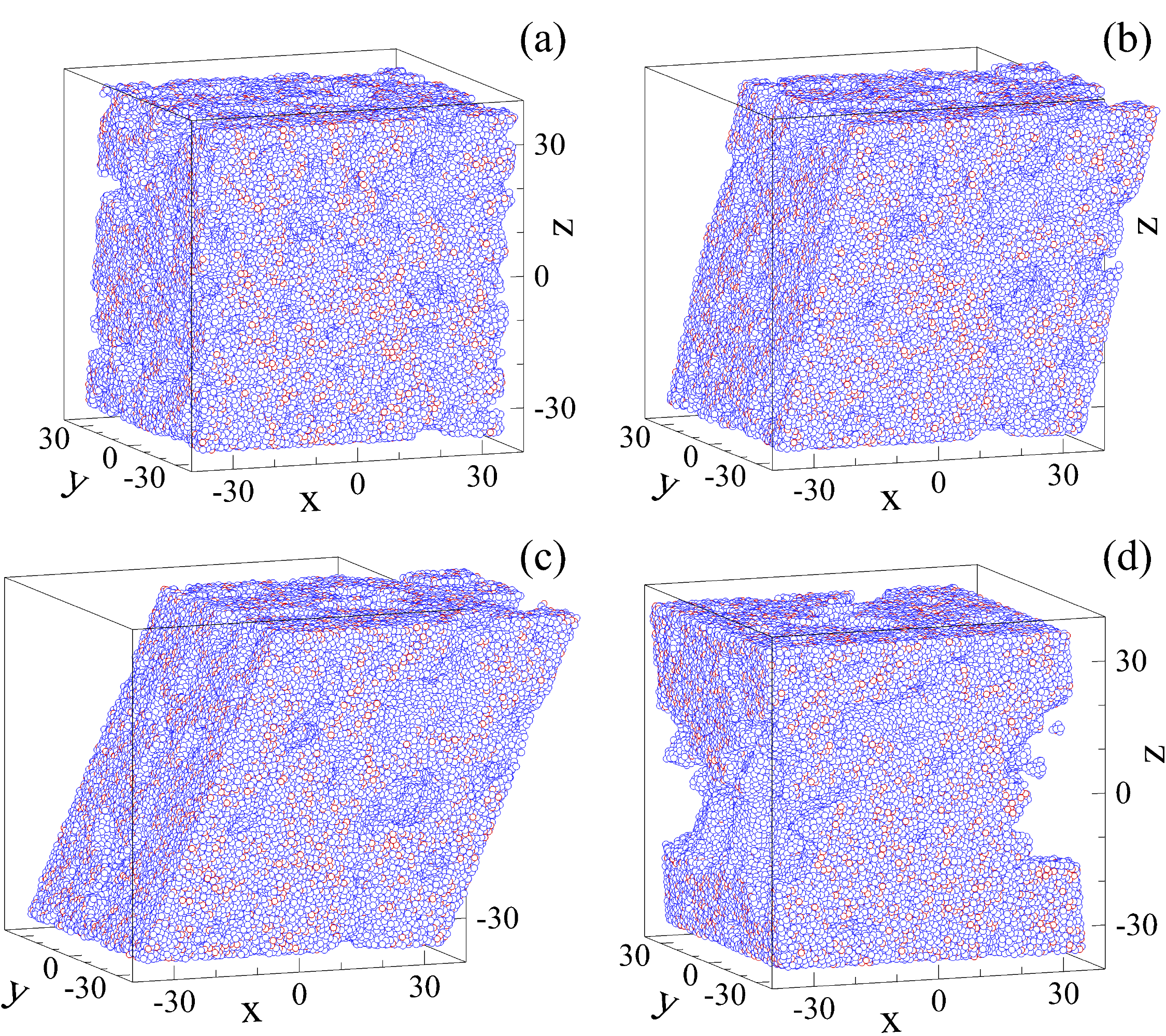}
\caption{(Color online) Spatial configurations of atoms for the
average glass density $\rho\sigma^{3}=0.8$ and strain (a)
$\gamma=0.05$, (b) $\gamma=0.25$, (c) $\gamma=0.45$, and (d)
$\gamma=0.95$. The same sample at zero strain is shown in
Fig.\,\ref{fig:snapshot_system}\,(d). }
\label{fig:snapshot_shear_rho08}
\end{figure}
\begin{figure}[t]
\includegraphics[width=15.cm,angle=0]{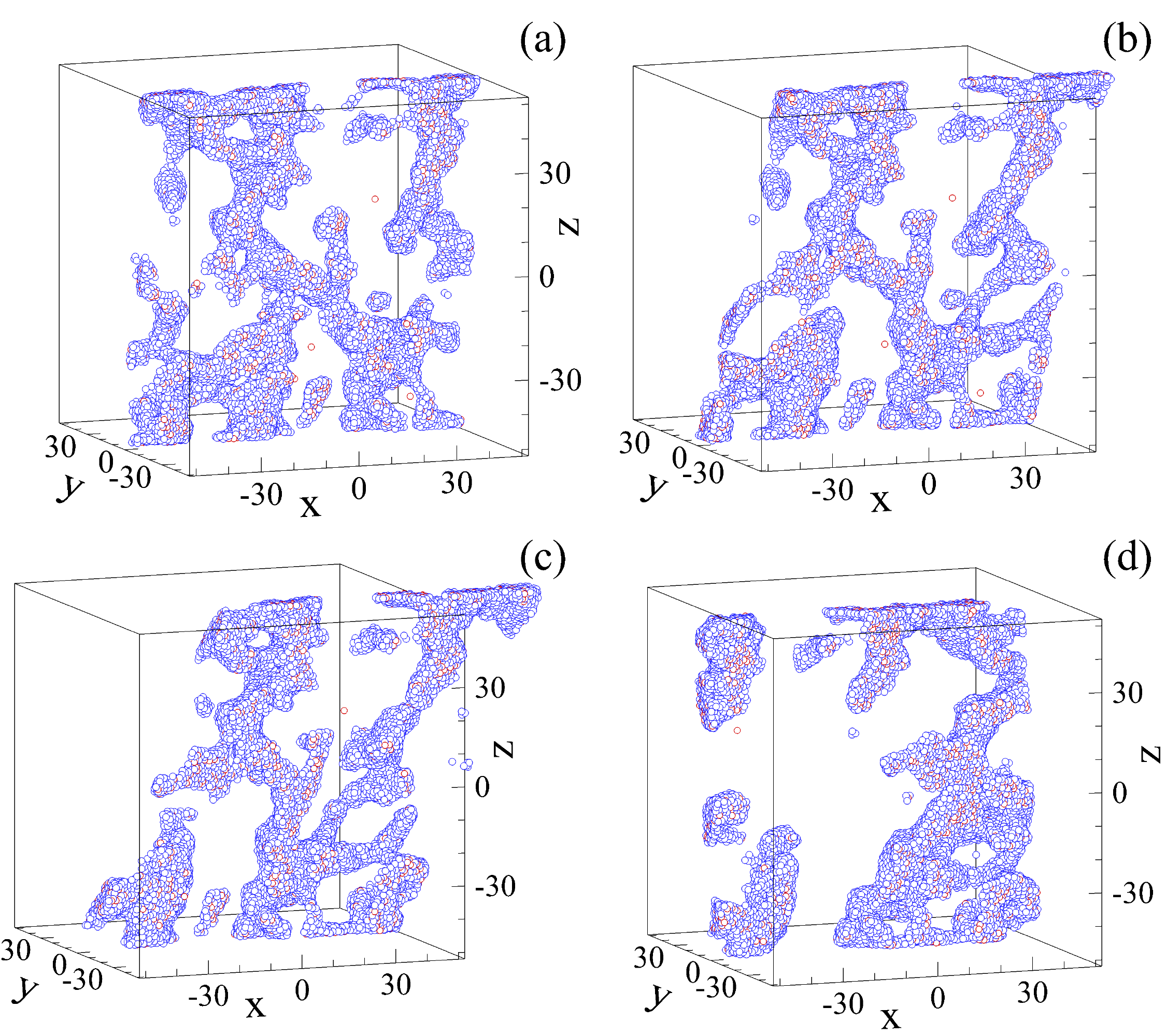}
\caption{(Color online) A sequence of atom configurations within a
slice of thickness $10\,\sigma$ for the average glass density
$\rho\sigma^{3}=0.3$ and shear strain (a) $\gamma=0.05$, (b)
$\gamma=0.25$, (c) $\gamma=0.45$, and (d) $\gamma=0.95$.   The same
data as in Fig.\,\ref{fig:snapshot_shear_rho03}.  }
\label{fig:snapshot_shear_rho03_slice}
\end{figure}
\begin{figure}[t]
\includegraphics[width=15.cm,angle=0]{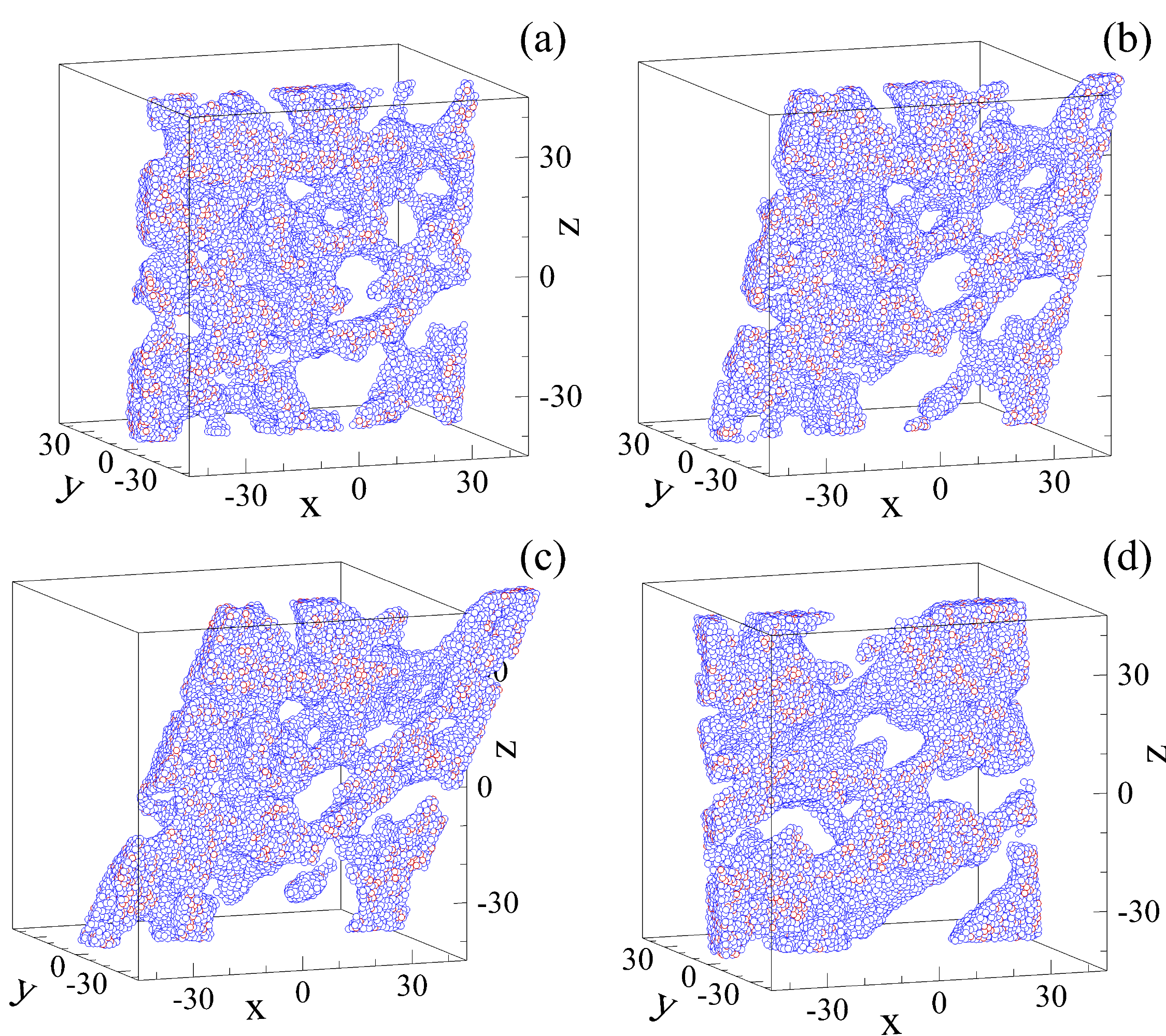}
\caption{(Color online) Snapshots of the sheared system within the
narrow slice of $10\,\sigma$ for the average glass density
$\rho\sigma^{3}=0.5$ and shear strain (a) $\gamma=0.05$, (b)
$\gamma=0.25$, (c) $\gamma=0.45$, and (d) $\gamma=0.95$. The same
data as in Fig.\,\ref{fig:snapshot_shear_rho05}. }
\label{fig:snapshot_shear_rho05_slice}
\end{figure}
\begin{figure}[t]
\includegraphics[width=15.cm,angle=0]{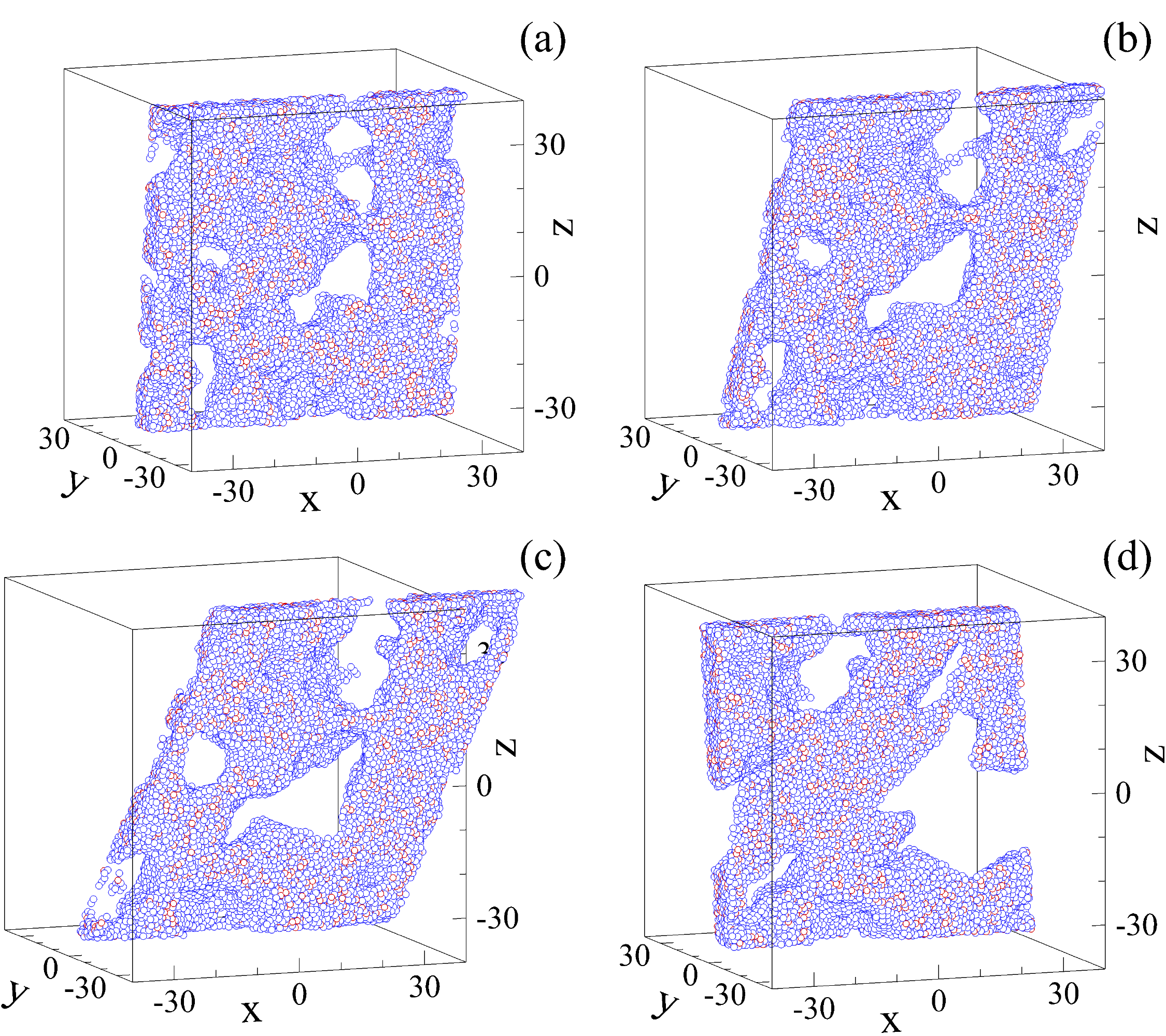}
\caption{(Color online) Spatial configurations of atoms in a thin
slice of $10\,\sigma$ for the average glass density
$\rho\sigma^{3}=0.8$ and strain (a) $\gamma=0.05$, (b)
$\gamma=0.25$, (c) $\gamma=0.45$, and (d) $\gamma=0.95$. The same
sample at zero strain is shown in
Fig.\,\ref{fig:snapshot_system}\,(d). The same data as in
Fig.\,\ref{fig:snapshot_shear_rho08}.   }
\label{fig:snapshot_shear_rho08_slice}
\end{figure}

The representative snapshots of atom configurations during steady
shear deformation are shown in
Figs.\,\ref{fig:snapshot_shear_rho03},
\ref{fig:snapshot_shear_rho05} and \ref{fig:snapshot_shear_rho08}
for the average glass densities $\rho\sigma^{3}=0.3$, $0.5$, and
$0.8$. To avoid highly skewed simulation box,  the configurations of
atoms in the case $\gamma=0.95$ are plotted within a tilted box with
the strain $\gamma=-0.05$.   From the consecutive snapshots, it can
be observed that pores become significantly deformed and, with
increasing shear strain, they tend to coalesce into larger voids.
The deformation process of individual pores during strain can be
more clearly visualized by plotting atom positions in a thin slice
of $10\,\sigma$ along the plane of shear, as shown in
Figs.\,\ref{fig:snapshot_shear_rho03_slice},
\ref{fig:snapshot_shear_rho05_slice} and
\ref{fig:snapshot_shear_rho08_slice}. Generally, the formation of
extended solid or void structures in a deformed glass is reflected
in a large deviation of shear stress from its averaged value.  For
example, notice the appearance of a large pore in the center of the
sample in Fig.\,\ref{fig:snapshot_shear_rho08}\,(d) and
Fig.\,\ref{fig:snapshot_shear_rho08_slice}\,(d) that correlates well
with the reduced value of shear stress for $\gamma \gtrsim 0.5$ and
$\rho\sigma^{3}=0.8$ in Fig.\,\ref{fig:stress_strain}.

\vskip 0.05in

\begin{figure}[t]
\includegraphics[width=12.cm,angle=0]{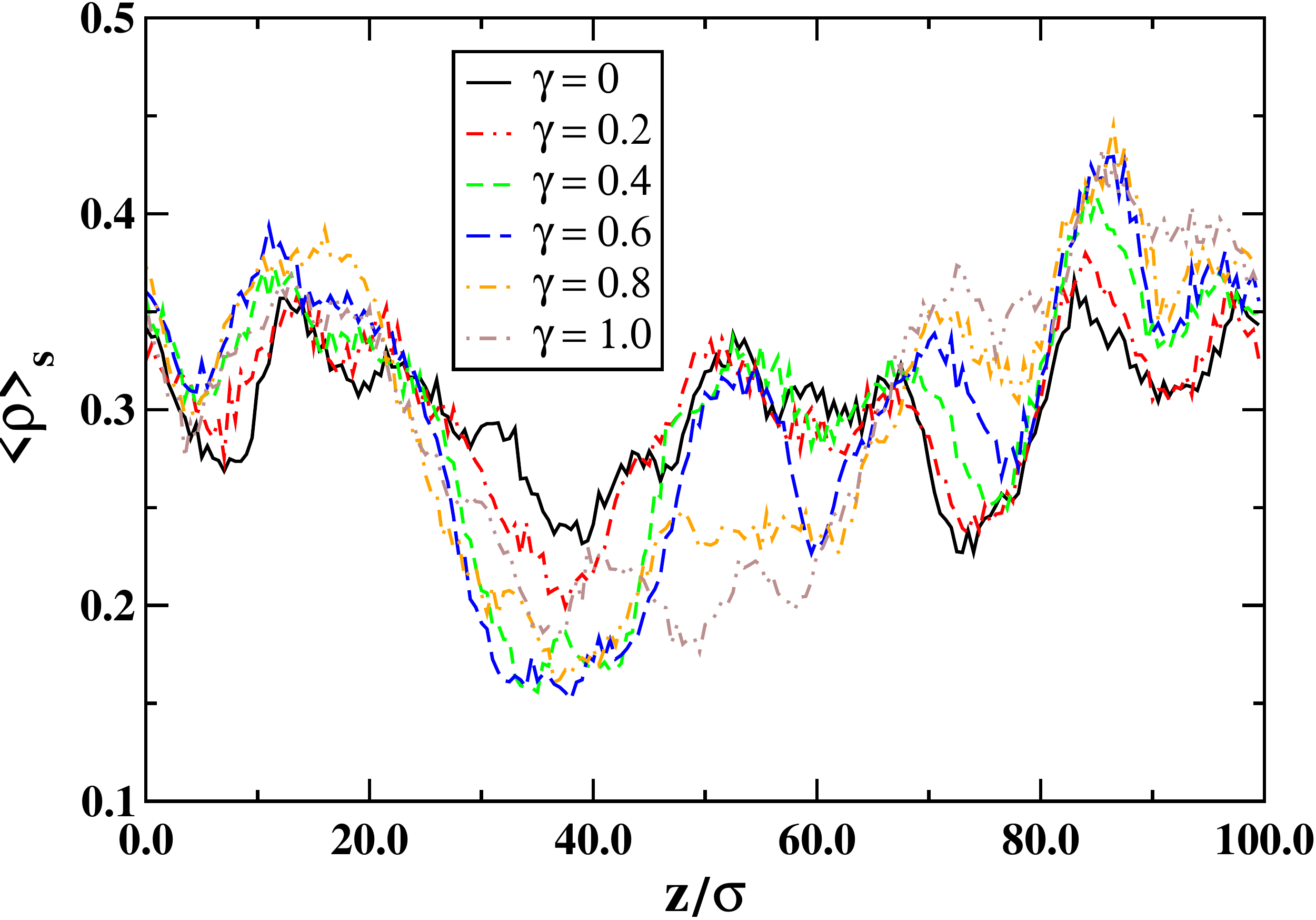}
\caption{(Color online) The density profiles $\langle \rho
\rangle_s$ (in units of $\sigma^{-3}$) along the $\hat{z}$ direction
for the indicated values of strain.  The data were averaged in thin
slices parallel to the $xy$ plane.  The average glass density is
$\rho\sigma^{3}=0.3$.}
\label{fig:den_prof_rho03}
\end{figure}
\begin{figure}[t]
\includegraphics[width=12.cm,angle=0]{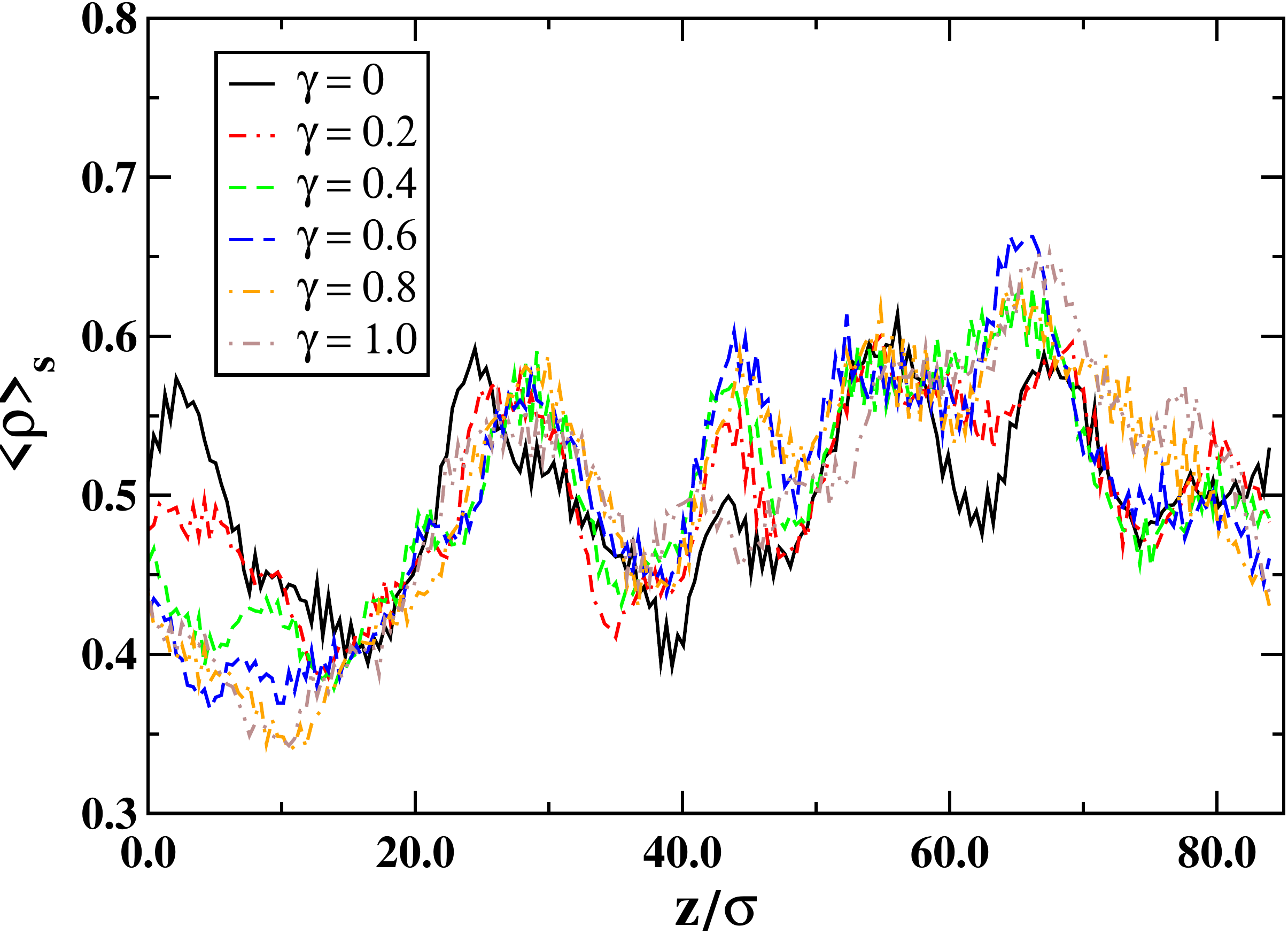}
\caption{(Color online) The averaged density profiles $\langle \rho
\rangle_{s}(z)$ (in units of $\sigma^{-3}$) for the selected values
of shear strain.  The average glass density is
$\rho\sigma^{3}=0.5$.}
\label{fig:den_prof_rho05}
\end{figure}
\begin{figure}[t]
\includegraphics[width=12.cm,angle=0]{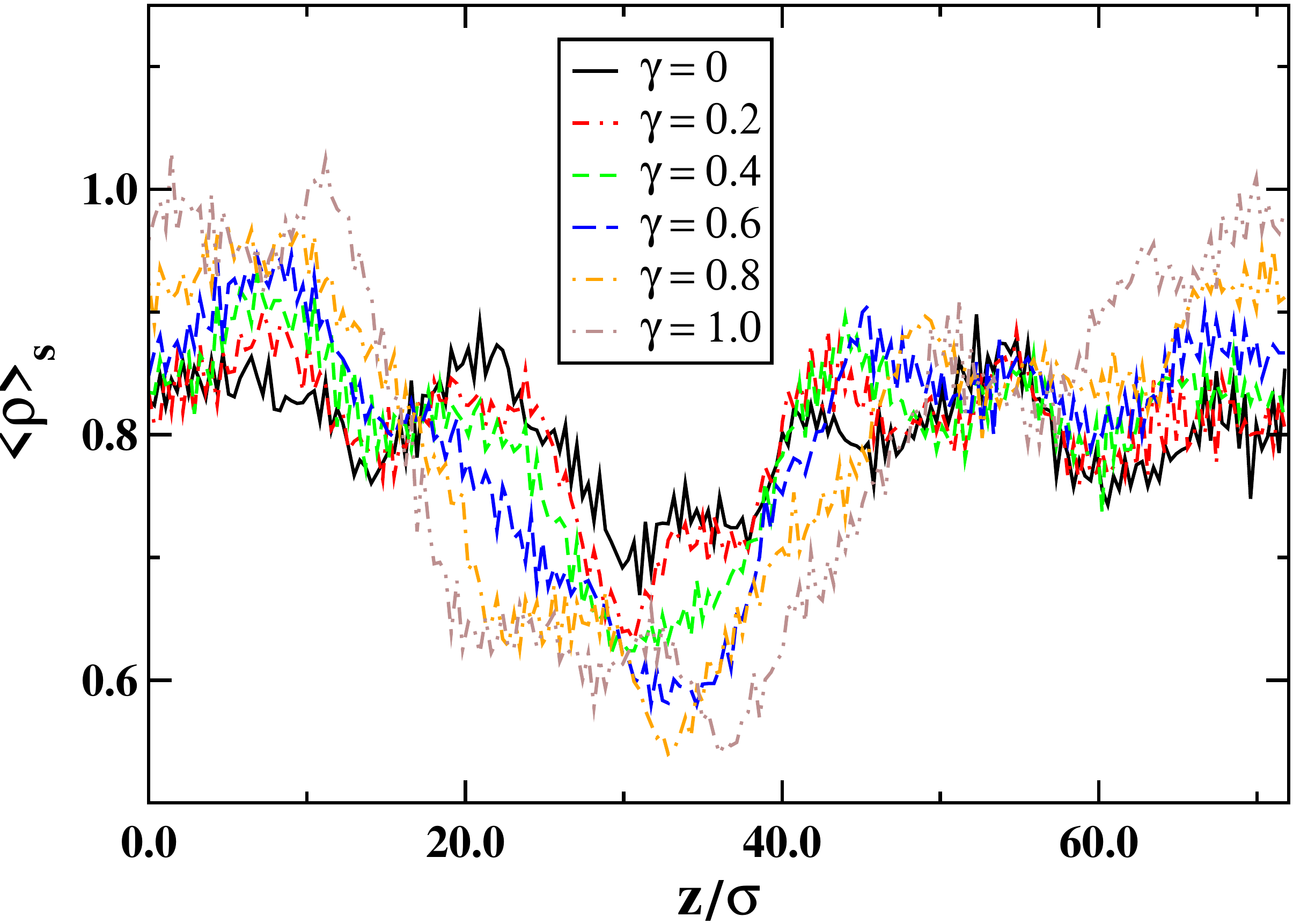}
\caption{(Color online) The atomic density profiles $\langle \rho
\rangle_{s}(z)$ (in units of $\sigma^{-3}$) as a function of shear
strain. The average glass density is $\rho\sigma^{3}=0.8$.}
\label{fig:den_prof_rho08}
\end{figure}

The position-dependent density profiles along the $\hat{z}$
direction are presented in Figs.\,\ref{fig:den_prof_rho03},
\ref{fig:den_prof_rho05} and \ref{fig:den_prof_rho08} for the
average glass densities $\rho\sigma^{3}=0.3$, $0.5$ and $0.8$,
respectively. In each case, the data were averaged in one sample at
different strain values in thin slices parallel to the $xy$ plane.
It can be clearly observed that the local density in quiescent
samples might be very different from the average bulk density.  By
definition, the local minima in density profiles correspond to one
or more pores that are not necessarily connected with each other.
Interestingly, it can also be noticed that with increasing shear
strain, one of the local minima is further reduced.   For example,
the following regions become less dense: $25\,\sigma\lesssim
z\lesssim45\,\sigma$ in Fig.\,\ref{fig:snapshot_shear_rho03_slice},
$5\,\sigma\lesssim z\lesssim20\,\sigma$ in
Fig.\,\ref{fig:snapshot_shear_rho05_slice}, and $20\,\sigma\lesssim
z\lesssim45\,\sigma$ in Fig.\,\ref{fig:snapshot_shear_rho08_slice}.
Thus, the shear-induced local expulsion of the solid phase suggests
the formation of larger pores and flow localization in narrow
regions. This behavior is consistent with the formation of shear
bands running across the system in strained homogeneous amorphous
materials~\cite{FalkPRL05,Gendel15,Horbach16,Priezjev17}.

\vskip 0.05in


In this study, the pore size distribution (PSD) functions were
computed using the Zeo++ software
tool~\cite{Haranczyk12c,Haranczyk12}. Correspondingly, the analysis
of pore-size distributions, employed herein, is fully based on the
algorithms and computational methods, described in
Refs.\,\cite{Haranczyk12c,Haranczyk12}. In brief, the analysis is
based on decomposition of each simulation system volume into Voronoi
cells, associated with each individual atom in the atomic system
under consideration. Thereby, a periodic Voronoi network is obtained
for each system (note that periodic boundary conditions are applied
in each case). The networks contain information on the nodes and
edges of the Voronoi cells. The code tests whether the total volume
of the systems is equal to the sum of the volumes of the computed
Voronoi cells. Each node and edge is labeled with its distance to
the corresponding set of nearest atoms. Thus, the obtained Voronoi
network represents the void space in a porous material. By defining
the probe radius, one can identify the probe-accessible regions of
the Voronoi network. The implementation of the method, used to
identify probe-accessible domains, is based on a variation of the
Dijkstra's shortest-path algorithm~\cite{Dijkstra59}. The PSDs are
computed using probe radius of $1.2\,\sigma$.

\vskip 0.05in

\begin{figure}[t]
\includegraphics[width=12.cm,angle=0]{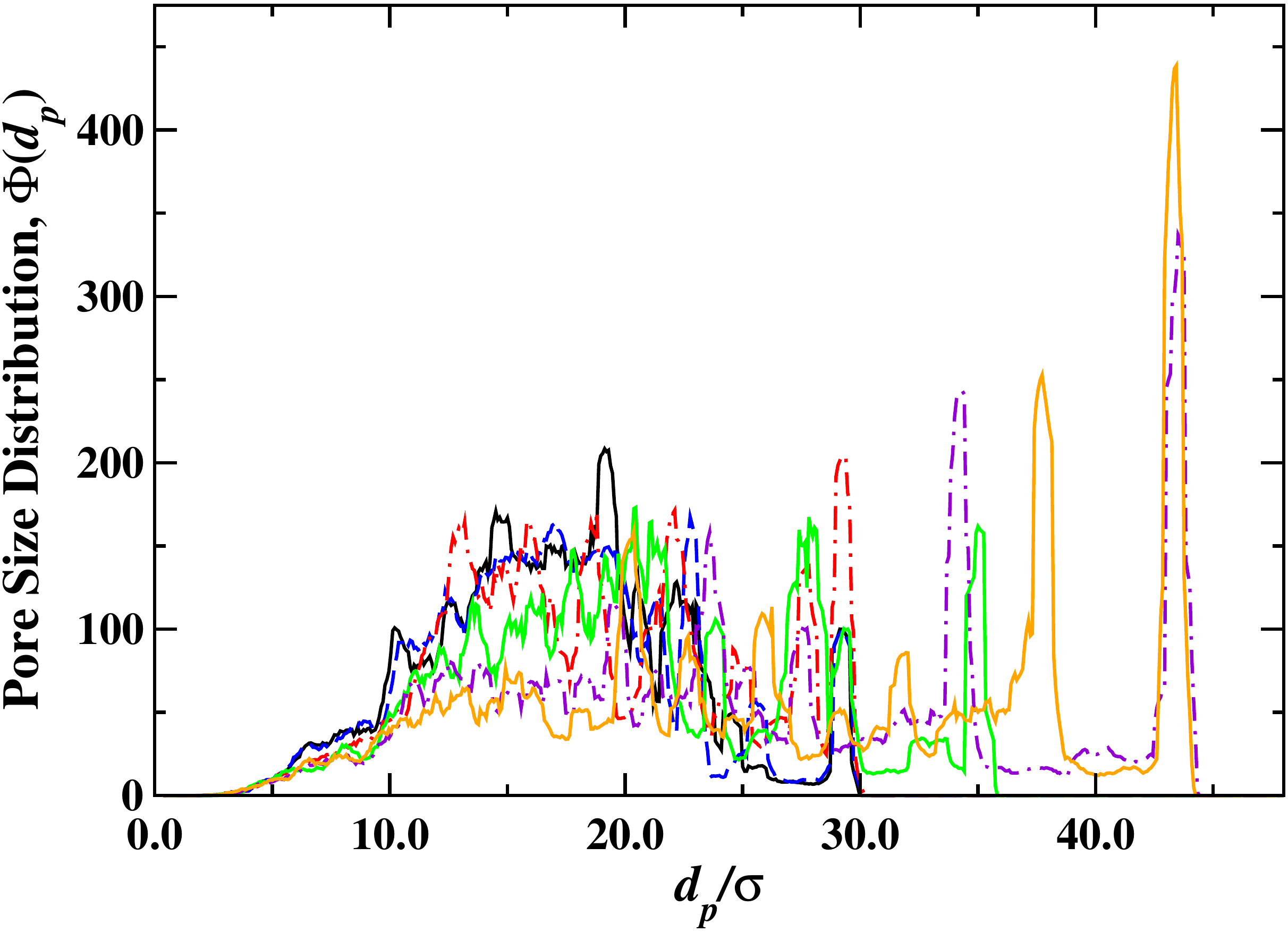}
\caption{(Color online) The distribution of pore sizes for the
average glass density $\rho\sigma^{3}=0.3$ and shear strain
$\gamma=0.0$ (black solid curve), $0.05$ (blue dashed curve), $0.25$
(red dash-dotted curve), $0.45$ (green solid curve), $0.75$ (velvet
dash-dotted curve), and $0.90$ (orange solid curve). }
\label{fig:pore_size_dist_rho03}
\end{figure}
\begin{figure}[t]
\includegraphics[width=12.cm,angle=0]{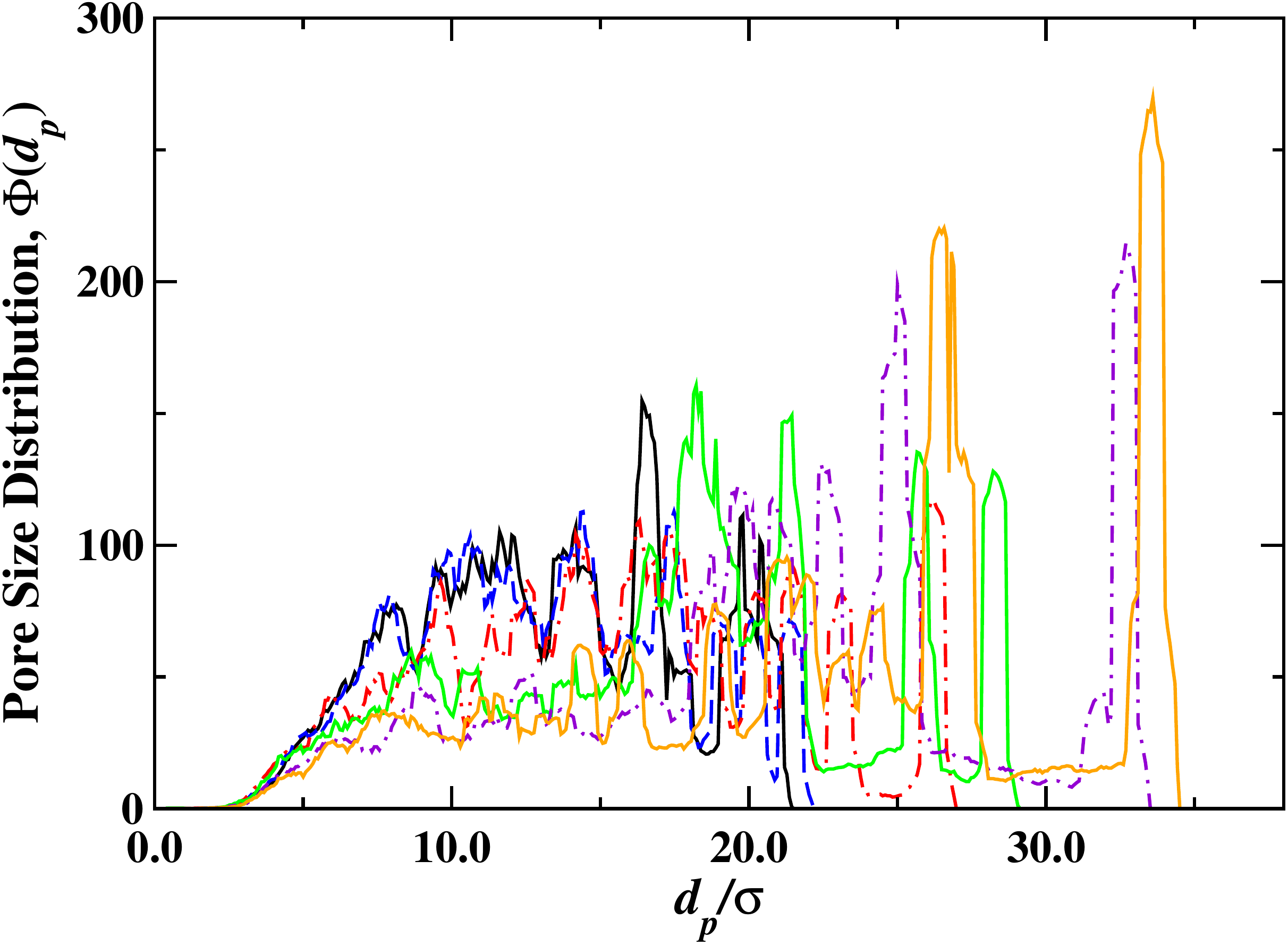}
\caption{(Color online) The pore size distribution for the density
$\rho\sigma^{3}=0.5$ and shear strain $\gamma=0.0$ (black solid
curve), $0.05$ (blue dashed curve), $0.25$ (red dash-dotted curve),
$0.45$ (green solid curve), $0.75$ (velvet dash-dotted curve), and
$0.90$ (orange solid curve).}
\label{fig:pore_size_dist_rho05}
\end{figure}
\begin{figure}[t]
\includegraphics[width=12.cm,angle=0]{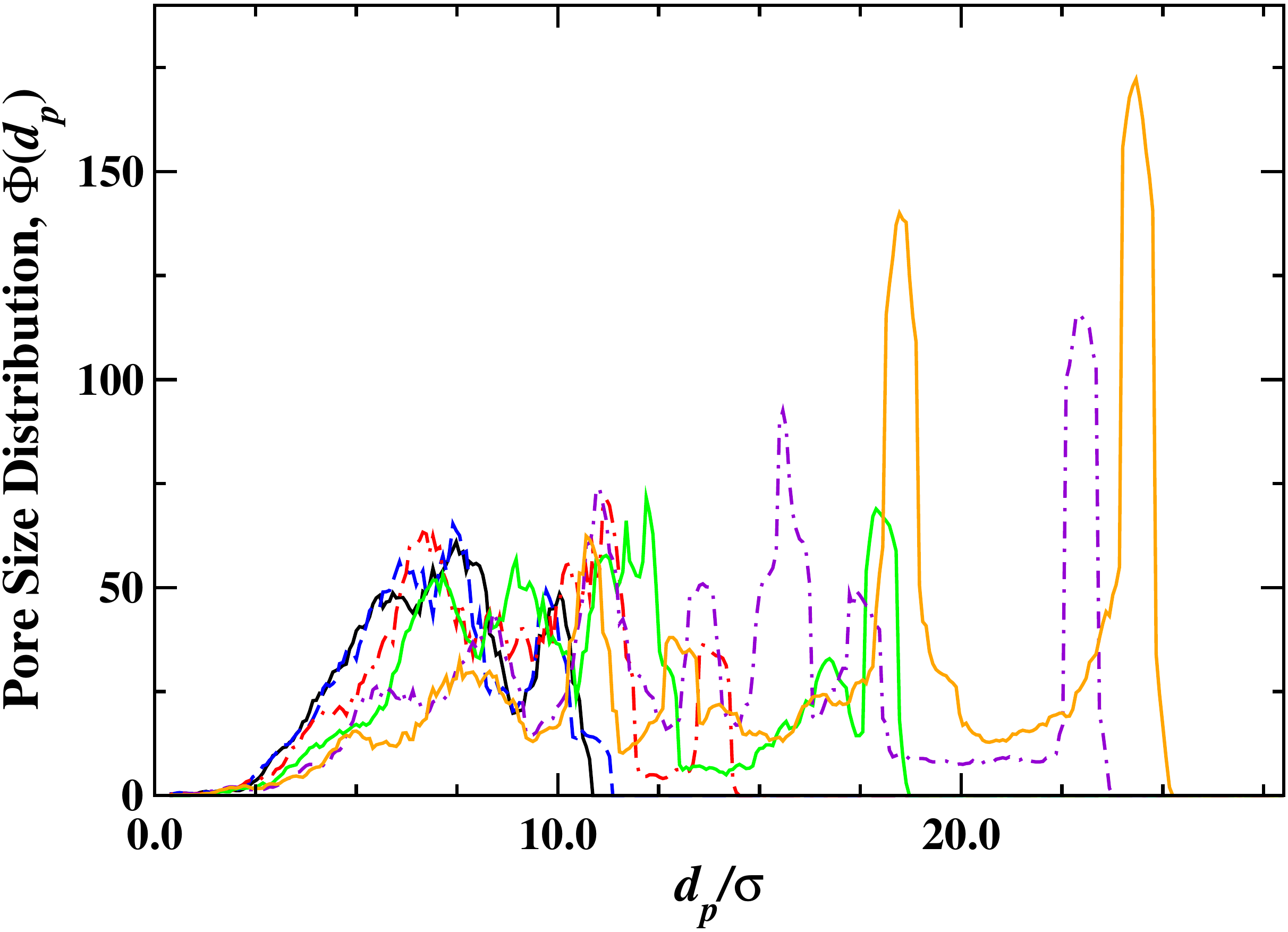}
\caption{(Color online) The probability distribution of pore sizes
for the average glass density $\rho\sigma^{3}=0.8$ and shear strain
$\gamma=0.0$ (black solid curve), $0.05$ (blue dashed curve), $0.25$
(red dash-dotted curve), $0.45$ (green solid curve), $0.75$ (velvet
dash-dotted curve), and $0.90$ (orange solid curve). }
\label{fig:pore_size_dist_rho08}
\end{figure}

The pore size distribution functions for sheared samples with
densities $\rho\sigma^{3}=0.3$, $0.5$ and $0.8$ are presented in
Figs.\,\ref{fig:pore_size_dist_rho03},
\ref{fig:pore_size_dist_rho05} and \ref{fig:pore_size_dist_rho08},
respectively.   In agreement with the previous MD
study~\cite{Makeev17}, the distribution of pore sizes in quiescent
samples is narrow at high glass densities and it becomes broader as
the average glass density decreases.   Under small shear
deformation, $\gamma=0.05$, the shape of PSD curves remains largely
unaffected, indicating that pores are only slightly deformed under
shear but do not coalesce. With further increasing strain, the
distribution functions are progressively skewed toward large values
of $d_p$, and a number of discrete peaks are developed at large
length scales. Note that the pronounced peaks in PSDs at large
length scales, for example, at $d_p\approx43.4\,\sigma$ for
$\gamma=0.90$ in Fig.\,\ref{fig:pore_size_dist_rho03}, signify the
appearance of elongated (`ellipsoid-like') porous structures with
the typical smallest size of about $d_p$.  In turn, the height of
peaks reflects the number of attempts that were used to identify
numerically the probe-accessible regions of the Voronoi network
associated with large pores. We further comment that the largest
length scale of elongated pores or the pore connectivity can not be
deduced from the PSD curves presented in
Figs.\,\ref{fig:pore_size_dist_rho03},
\ref{fig:pore_size_dist_rho05} and \ref{fig:pore_size_dist_rho08}.
Altogether, it can be concluded that shear deformation is
accompanied with a structural transition from a number of compact
pores to a configuration with one or two dominant pores.  This
interpretation is consistent with the visual inspection of deformed
samples presented in Figs.\,\ref{fig:snapshot_shear_rho03},
\ref{fig:snapshot_shear_rho05} and \ref{fig:snapshot_shear_rho08}
where the formation of large pores is evident at large strain.

\section{Conclusions}

In summary, molecular dynamics simulations were performed to examine
mechanical properties and structural morphology of porous glasses
under steady shear.   We considered the three-dimensional binary
mixture that was quenched at constant volume from a high-temperature
liquid phase to a temperature well below glass transition.  The
coarsening process involved formation of disordered structures that
were characterized by a relatively narrow pore size distribution
with the largest length scale that depends on the average glass
density.  It was found that at small shear strain, the pore network
becomes linearly transformed while its connectivity remains
unchanged. Moreover, the dependence of shear modulus on porosity at
low glass densities agrees well with continuum predictions based on
the percolation theory.  A strikingly different behavior occurs at
large strain when the pore shape becomes extremely distorted and
adjacent pores coalesce into large voids, which is reflected in the
development of a skewed pore size distribution with a superimposed
discrete peaks. Thus, the size of the largest pore is greater at
lower average glass densities and higher shear strain.

\section*{Acknowledgments}

Financial support from the National Science Foundation (CNS-1531923)
is gratefully acknowledged.  The authors would like to thank
Dr.~M.~Haranczyk (Lawrence Berkeley National Laboratory) for his
help with the Zeo++ code.   The molecular dynamics simulations were
carried out using the LAMMPS numerical code developed at Sandia
National Laboratories~\cite{Lammps}. This work was supported in part
by Michigan State University through computational resources
provided by the Institute for Cyber-Enabled Research.


%

%

%

%

%

%

%

%

%

%

%

%

%

%

%

%

\bibliographystyle{prsty}

\end{document}